\documentclass{moriond}

\def\be{\begin{equation}}
\def\ee{\end{equation}}
\def\bea{\begin{eqnarray}}
\def\eea{\end{eqnarray}}

\newcommand{\Photo}{\vspace{+0.1cm} \includegraphics[height=35mm]{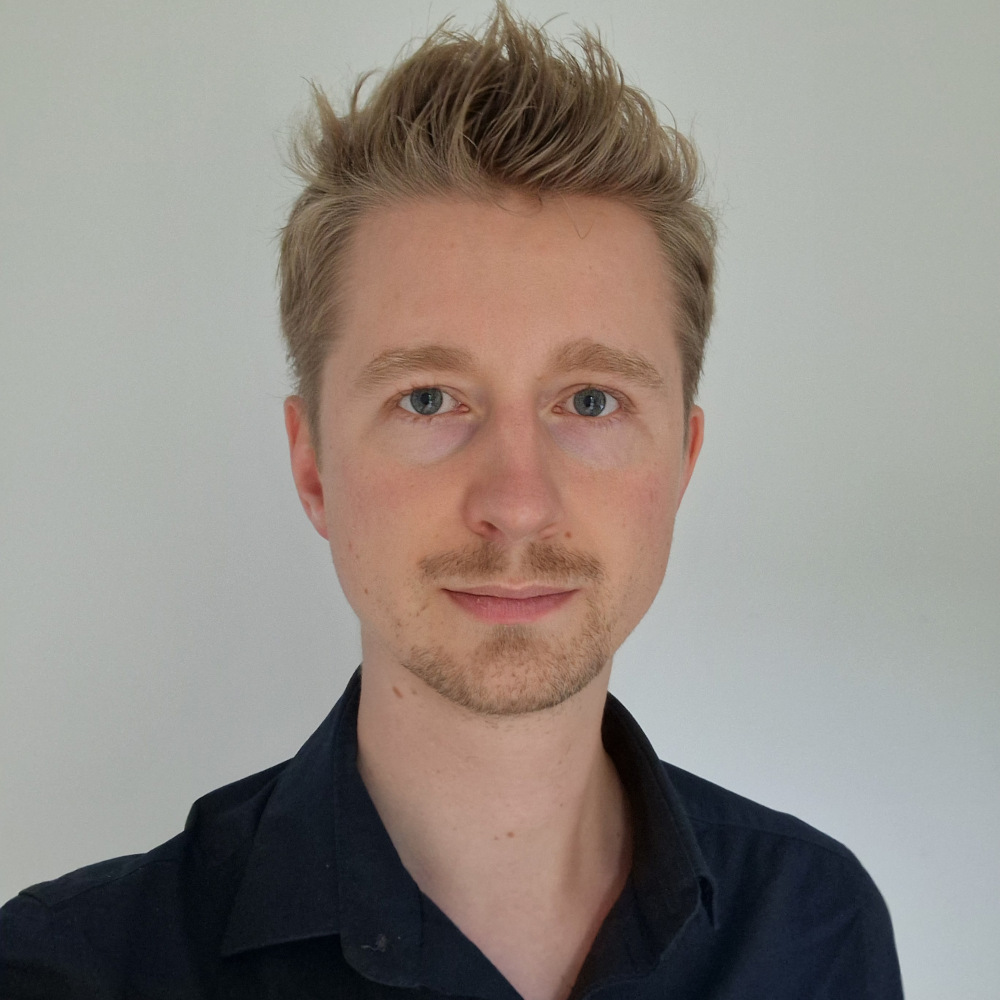}\vspace{-0.1cm}}

\usepackage{xspace} 
\usepackage{upgreek}

\def\MagUp {\mbox{\em Mag\kern -0.05em Up}\xspace}

\def\Pmu         {\ensuremath{\mu}\xspace}

\def\Ppi         {\ensuremath{\pi}\xspace}

\def\Ppsi        {\ensuremath{\psi}\xspace}                 
                 
\mathchardef\PDelta="7101
\mathchardef\PXi="7104
\mathchardef\PLambda="7103
\mathchardef\PSigma="7106
\mathchardef\POmega="710A
\mathchardef\PUpsilon="7107
                 
\def\PB      {\ensuremath{B}\xspace}                 
                 
\def\PD      {\ensuremath{D}\xspace}

\def\PJ      {\ensuremath{J}\xspace}                 
\def\PK      {\ensuremath{K}\xspace}

\def\Pd      {\ensuremath{d}\xspace}

\def\Pi      {\ensuremath{i}\xspace}

\def\Ps      {\ensuremath{s}\xspace}

\def\mup        {{\ensuremath{\Pmu^+}}\xspace}
\def\mun        {{\ensuremath{\Pmu^-}}\xspace} 

\def\ellm       {{\ensuremath{\ell^-}}\xspace}
\def\ellp       {{\ensuremath{\ell^+}}\xspace}

\def\dquark    {{\ensuremath{\Pd}}\xspace}

\def\squark    {{\ensuremath{\Ps}}\xspace}

\def\pion   {{\ensuremath{\Ppi}}\xspace}
\def\piz    {{\ensuremath{\pion^0}}\xspace}

\def\pip    {{\ensuremath{\pion^+}}\xspace}
\def\pim    {{\ensuremath{\pion^-}}\xspace}

\def\pimp   {{\ensuremath{\pion^\mp}}\xspace}

\def\kaon    {{\ensuremath{\PK}}\xspace}
\def\Kbar    {{\kern 0.2em\overline{\kern -0.2em \PK}{}}\xspace}

\def\KorKbar    {\kern 0.18em\optbar{\kern -0.18em K}{}\xspace}

\def\Kp      {{\ensuremath{\kaon^+}}\xspace}
\def\Km      {{\ensuremath{\kaon^-}}\xspace}
\def\Kpm     {{\ensuremath{\kaon^\pm}}\xspace}

\def\KS      {{\ensuremath{\kaon^0_{\mathrm{ \scriptscriptstyle S}}}}\xspace}

\def\Dbar    {{\kern 0.2em\overline{\kern -0.2em \PD}{}}\xspace}
\def\D       {{\ensuremath{\PD}}\xspace}

\def\DorDbar    {\kern 0.18em\optbar{\kern -0.18em D}{}\xspace}
\def\Dz      {{\ensuremath{\D^0}}\xspace}
\def\Dzb     {{\ensuremath{\Dbar{}^0}}\xspace}

\def\Dstarp  {{\ensuremath{\D^{*+}}}\xspace}

\def\Dsp     {{\ensuremath{\D^+_\squark}}\xspace}
\def\Dsm     {{\ensuremath{\D^-_\squark}}\xspace}

\def\B       {{\ensuremath{\PB}}\xspace}
\def\Bbar    {{\ensuremath{\kern 0.18em\overline{\kern -0.18em \PB}{}}}\xspace}

\def\BorBbar {\kern \thebaroffset\optbar{\kern -\thebaroffset \PB}\xspace}

\def\Bd      {{\ensuremath{\B^0}}\xspace}
\def\Bdb     {{\ensuremath{\Bbar{}^0}}\xspace}
\def\BdorBdbar {\kern \thebaroffset\optbar{\kern -\thebaroffset \Bd}\xspace}
\def\Bu      {{\ensuremath{\B^+}}\xspace}
\def\Bub     {{\ensuremath{\B^-}}\xspace}
\def\Bp      {{\ensuremath{\Bu}}\xspace}
\def\Bm      {{\ensuremath{\Bub}}\xspace}
\def\Bpm     {{\ensuremath{\B^\pm}}\xspace}

\def\Bs      {{\ensuremath{\B^0_\squark}}\xspace}
\def\Bsb     {{\ensuremath{\Bbar{}^0_\squark}}\xspace}
\def\BsorBsbar {\kern \thebaroffset\optbar{\kern -\thebaroffset \Bs}\xspace}

\def\jpsi     {{\ensuremath{{\PJ\mskip -3mu/\mskip -2mu\Ppsi}}}\xspace}

\def\Y#1S{\ensuremath{\PUpsilon{(#1S)}}\xspace}

\def\Lbar        {{\ensuremath{\kern 0.1em\overline{\kern -0.1em\PLambda}}}\xspace}
\def\LorLbar    {\kern 0.18em\optbar{\kern -0.18em \PLambda}{}\xspace}

\newcommand{\decay}[2]{\ensuremath{#1\!\to #2}\xspace}         

\def\to                 {\ensuremath{\rightarrow}\xspace}

\newcommand{\tauDz}{{\ensuremath{\tau_{\Dz}}}\xspace}

\def\CP                {{\ensuremath{C\!P}}\xspace}

\newcommand{\DGs}{{\ensuremath{\Delta\Gamma_{\squark}}}\xspace}

\newcommand{\Gs}{{\ensuremath{\Gamma_{\squark}}}\xspace}
\newcommand{\Gd}{{\ensuremath{\Gamma_{\dquark}}}\xspace}

\def\AT#1     {\ensuremath{A_{\mathrm{T}}^{#1}}\xspace}           

\def\C#1      {\ensuremath{\mathcal{C}_{#1}}\xspace}                       
\def\Cp#1     {\ensuremath{\mathcal{C}_{#1}^{'}}\xspace}                    
\def\Ceff#1   {\ensuremath{\mathcal{C}_{#1}^{\mathrm{(eff)}}}\xspace}        
\def\Cpeff#1  {\ensuremath{\mathcal{C}_{#1}^{'\mathrm{(eff)}}}\xspace}       
\def\Ope#1    {\ensuremath{\mathcal{O}_{#1}}\xspace}                       
\def\Opep#1   {\ensuremath{\mathcal{O}_{#1}^{'}}\xspace}                    

\def\invfb   {\ensuremath{\mbox{\,fb}^{-1}}\xspace}

\def\ps   {\ensuremath{{\mathrm{ \,ps}}}\xspace}

\newcommand{\chisq}{\ensuremath{\chi^2}\xspace}

\def\gsim{{~\raise.15em\hbox{$>$}\kern-.85em
		\lower.35em\hbox{$\sim$}~}\xspace}
\def\lsim{{~\raise.15em\hbox{$<$}\kern-.85em
		\lower.35em\hbox{$\sim$}~}\xspace}

\def\degrees{\ensuremath{^{\circ}}\xspace}

\def\rad{\ensuremath{\mathrm{ \,rad}}\xspace}

\def\tell1  {TELL1\xspace}
\def\ukl1   {UKL1\xspace}

\newcommand{\rdsk}      {\ensuremath{r_{D_sK}}\xspace}

\newcommand{\BsDsK}    {\texorpdfstring{\decay{B^0_s}{D_s^\mp K^\pm}}{}}
\newcommand{\BsDsPi}   {\texorpdfstring{\decay{B^0_s}{D_s^- \pi^+}}{}}

\def\tauDz {\ensuremath{\tau_{\Dz}}\xspace}

\def\RKpip {\ensuremath{R_{\kaon\pion}^{+}}\xspace}
\def\RKpim {\ensuremath{R_{\kaon\pion}^{-}}\xspace}
\def\RKpipm {\ensuremath{R_{\kaon\pion}^{\pm}}\xspace}

\def\RKpi {\ensuremath{R_{\kaon\pion}}\xspace}
\def\AKpi {\ensuremath{A_{\kaon\pion}}\xspace}
\def\cKpi {\ensuremath{c_{\kaon\pion}}\xspace}
\def\cKpipr {\ensuremath{c_{\kaon\pion}^{\prime}}\xspace}
\def\DcKpi {\ensuremath{\Delta c_{\kaon\pion}}\xspace}
\def\DcKpipr {\ensuremath{\Delta c_{\kaon\pion}^{\prime}}\xspace}

\usepackage{amsmath}

\begin{document}
\vspace*{4cm}

\title{Mixing and \texorpdfstring{$\CP$}{CP} violation in beauty and charm at LHCb}

\author{Jordy Butter, on behalf of the LHCb Collaboration }

\address{Department of Physics, Cavendish Laboratory, JJ Thomson Avenue,\\
Cambridge, CB3 0HE, UK}
\maketitle
\abstracts{
These proceedings discuss recent measurements by the LHCb experiment on mixing and \mbox{$\CP$ violation} with beauty and charm mesons, as presented at the Moriond QCD 2024 conference.
All discussed measurements show agreement with the Standard Model. 
}

\section{Beauty measurements}
The study of $\B$ hadrons provides a great laboratory to study the CKM sector of the Standard Model (SM) and look for manifestations of physics beyond the SM.
First, direct determinations of the weak angle $\gamma$ using $\Bpm \rightarrow D^* h^{\pm}$ and $\Bd \rightarrow D K^*(892)^0$ decays are discussed.
    Afterwards, mixing-induced $\CP$ violation is studied using $\BsDsK$,  $\Bs \to \jpsi \Kp \Km$ and \mbox{$\Bd \to \psi(\to \ellp\ellm) \KS(\to \pip \pim)$} decays, probing the CKM angles $\gamma$, $\beta_s$ and $\beta$, respectively.
Finally, a measurement of $\DGs$ is shown.

\subsection{Direct determinations of the CKM angle \texorpdfstring{$\gamma$}{gamma}}
The experimental sensitivity to the CKM angle $\gamma$ is constrained most notably by $\Bpm \to D \Kpm$ decays, where $D$ is an admixture of $\Dz$ and $\Dzb$ mesons and the interference between $b\to c$ and $b\to u$ amplitudes is used to measure $\gamma$.
In these decays, the differences in the $D$-decay Dalitz-plot distributions between $\Bp$ and $\Bm$ decays are due to the $\CP$-violating phase $\gamma$, which is probed via time-integrated analyses.

These proceedings show two $\gamma$ measurements using $\Bpm \rightarrow D^* h^{\pm}$ decays, where $D^* \to D \gamma/\piz$ and $D\to \KS h^{+}h^{-}$, using proton-proton data collected between 2011 and 2012 (Run~1) and between 2015 and 2018 (Run~2), corresponding to an integrated luminosity of 3 and $6\invfb$, respectively.
The rich interference structure in $D\to \KS h^+ h^-$ decays, with $h={\pi, K}$, is explored through a model-independent approach, not relying on a $D$ decay amplitude model.
The interference effects are studied by measuring the yields of the $\Bp$ and $\Bm$ decays in bins of \mbox{$D$ decay} Dalitz phase space.
The key difference between these two measurements is that one fully reconstructs the final-state particles~\cite{LHCb-PAPER-2023-012}, whereas the other does not reconstruct the neutral $\gamma$ or $\piz$ coming from the $D^*$ decay~\cite{LHCb-PAPER-2023-029}.
The values of $\gamma = (69^{+13}_{-14})\degrees$ and $\gamma = (92^{+21}_{-17})\degrees$ are found for the fully and the partially reconstructed analysis, respectively.

\begin{figure}[t]
\vspace{-0.5\baselineskip}
\centering
\includegraphics[width=0.3\textwidth]{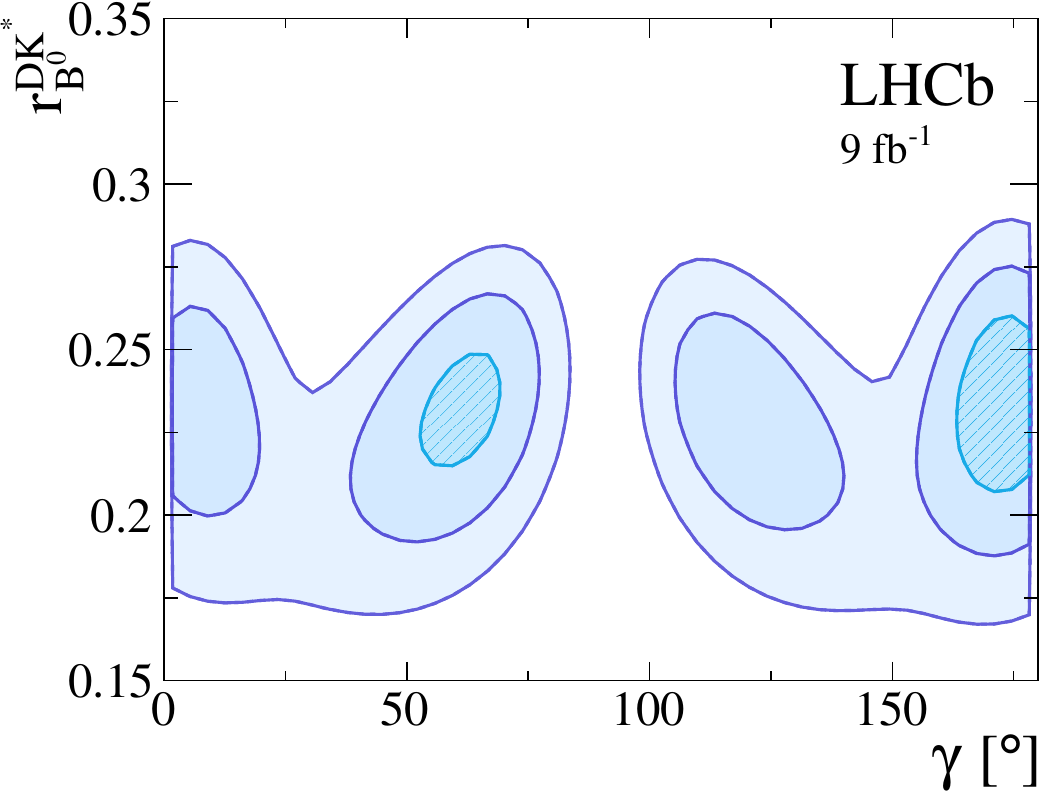}
\hspace{1.75cm}
\includegraphics[width=0.3\textwidth]{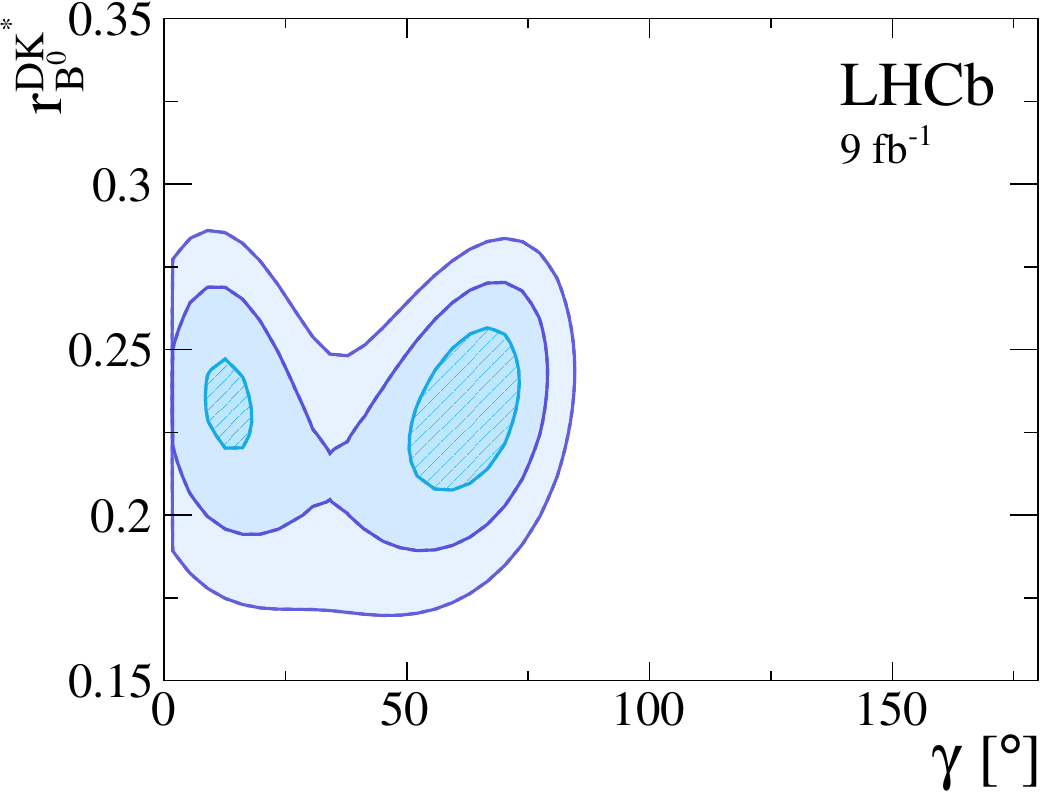}
\caption[]{Confidence-level contours projected in the $\gamma$--$r_{\Bd}^{DK^*}$ plane, obtained from $\Bd \rightarrow D K^*(892)^0$ decays with (left) $D\to K \pi (\pi\pi)$, $\pi\pi(\pi\pi)$ and $KK$ only and (right) combined with $D\to \KS h^+ h^-$. The contours contain $68.3\%$, $95.4\%$ and $99.7\%$ of the distribution.}
\label{fig:Bd2DKst_contours}
\end{figure}

Additionally, two measurements using $\Bd \rightarrow D K^*(892)^0$ decays are presented, using Run~1 and Run~2 data.
Despite its relatively small branching fraction, the study of the \mbox{$\Bd \rightarrow D K^*(892)^0$} decay provides a precise determination of $\gamma$ due to the relatively similar magnitude of the interfering amplitudes.
The first analysis studies interference effects using $D\to \KS h^+ h^-$ decays by measuring the $\Bd$ and $\Bdb$ decay yields across the $D$ phase space~\cite{LHCb-PAPER-2023-009}.
These yields are used to determine the weak phase $\gamma = (49^{+22}_{-19})\degrees$, the amplitude ratio $r_\Bd = 0.271^{+0.065}_{-0.066}$ and the strong phase $\delta_\Bd = (236^{+19}_{-21})\degrees$.
The other $\Bd \rightarrow D K^{*0}$ measurement studies final states with $D\to K \pi (\pi\pi)$, $\pi\pi(\pi\pi)$ and $KK$~\cite{LHCb-PAPER-2023-040}.
Invariant mass fits are used to measure asymmetries and ratios of the $\Bd$ and $\Bdb$ decay rates, which are functions of $\gamma$, $r_\Bd$ and $\delta_\Bd$.
Four solutions are found, but a combination with the previously described analysis favours the values $\gamma = (63.3\pm 7.2)\degrees$, $r_\Bd = 0.233 \pm 0.016$ and $\delta_\Bd = (191.9 \pm 6.0)\degrees$.
Figure~\ref{fig:Bd2DKst_contours} shows the confidence-level contours of the corresponding solutions projected in the $\gamma$--$r_{\Bd}^{DK^*}$ plane.

\subsection{Mixing-induced \texorpdfstring{$\CP$}{CP} violation}
In certain $\B$ hadron decays $\CP$ violation manifests itself in the interference between the mixed and the unmixed decay amplitudes.
This creates a time-dependent $\CP$ asymmetry, dependent on the complex phases of the CKM matrix.

\begin{figure}[t]
\vspace{-0.5\baselineskip}
\centering
\begin{minipage}{0.37\linewidth}
\includegraphics[width=\linewidth]{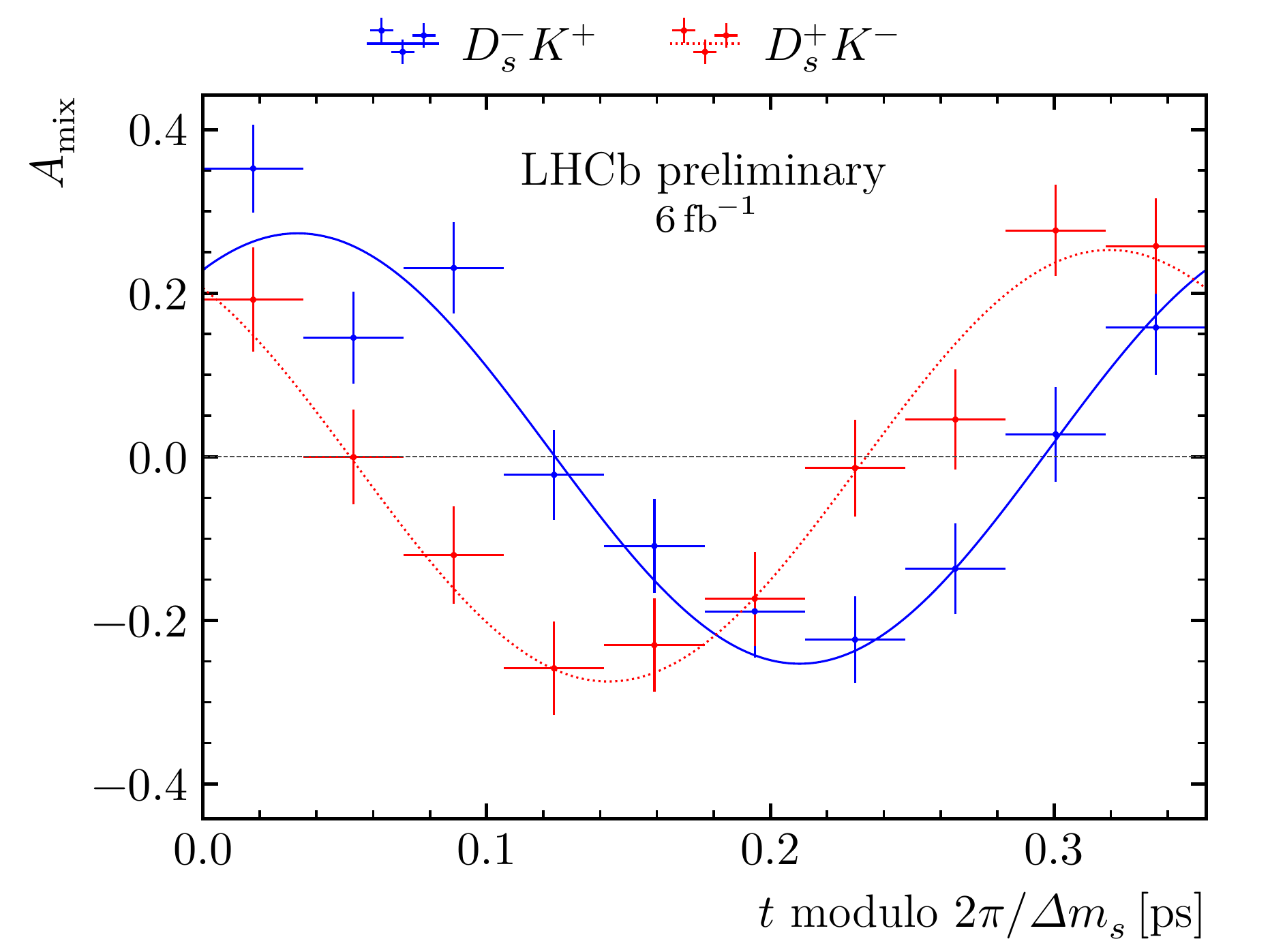}
\caption[]{
$\BsDsK$ mixing asymmetries of the (blue) $\Dsm\Kp$ (red) $\Dsp\Km$. The decay time is folded into a single oscillation period.
}
\label{fig:DsK_asymmetry}
\end{minipage}
\hfill
\begin{minipage}{0.39\linewidth}
\includegraphics[width=\linewidth]{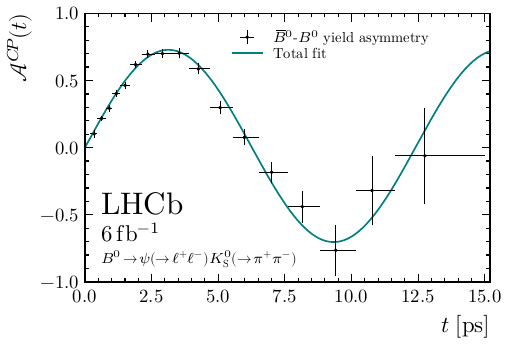}
\caption[]{
$\Bd \to \psi(\to \ellp\ellm) \KS(\to \pip \pim)$ time-dependent $\CP$-asymmetry
}
\label{fig:PsiKs_ACP}
\end{minipage}
\end{figure}

A well-known example of this type of $\CP$ violation is the study of $\BsDsK$ decays, where the interfering amplitudes have a comparable magnitude and a weak-phase difference of $\gamma - 2\beta_s$.
This analysis has recently been performed using Run~2 data~\cite{LHCb-CONF-2023-004} and makes use of the measurement of the $\Bs$--$\Bsb$ mixing frequency $\Delta m_s$ with $\BsDsPi$ decays to calibrate detector effects and for the value of $\Delta m_s$~\cite{LHCb-PAPER-2021-005}.
First, a likelihood fit is performed to the invariant mass of the $\Bs$ and $\Dsm$ mesons to separate the signal from the background.
Afterwards, a decay-time fit is performed to extract a set of five $\CP$ observables.
The corresponding $\CP$ asymmetry plot is provided in Fig.~\ref{fig:DsK_asymmetry}, where the shift between the two lines is a manifestation of $\CP$ violation.
The decay-time fit requires a good understanding of the decay-time acceptance, the resolution of the decay-time measurement and the flavour tagging, the main detector effects diluting the sensitivity to $\CP$ violation.
Finally, assuming $\phi_s= -2 \beta_s$ and using the LHCb combination of $\phi_s$~\cite{LHCb-PAPER-2023-016}, the measured $\CP$ observables are used to find the CKM angle $\gamma=(74\pm 11)\degrees$, the strong-phase difference $\delta = (346.9\pm6.6)^{\circ}$ and the amplitude ratio $\rdsk=0.327\pm0.038$.
This is the most precise determination of $\gamma$ with $\Bs$ mesons and agrees with the LHCb combination of $\gamma$~\cite{LHCb-CONF-2022-003}.

The most important constraint for the CKM angle $\beta_s$ is the study of $\Bs\to\jpsi\phi$ decays, where the measurable $\CP$-violating phase $\phi_s$ is equal to $-2\beta_s$ in the SM, neglecting subleading loop contributions.
The corresponding Run~2 measurement is performed using $\Bs\to\jpsi\Kp\Km$ decays near the $\phi(2010)$ resonance~\cite{LHCb-PAPER-2023-016}.
The measurement requires a flavour-tagged, time-dependent angular analysis, leading to the most precise determination of $\CP$ violating phase $\phi_s$, the $\Bs$ decay-width difference $\DGs$ and the difference in the average $\Bs$ and $\Bd$ decay width, $\Gs - \Gd$, to date.
This measurement finds $\phi_s = -0.039 \pm 0.022 \pm 0.006 \rad$,  $\DGs = 0.0845 \pm 0.0044 \pm 0.0024 \ps^{-1}$ and  $\Gs - \Gd = -0.0056^{+0.0013}_{-0.0015} \pm 0.0014 \ps^{-1}$, where the uncertainties are statistical and systematic.
The result is compatible with $\CP$ symmetry, other $\phi_s$ measurements and the SM expectation.

Also, the most precise single determination of the CKM angle $\beta$ is presented, using $\Bd \to \psi(\to \ellp\ellm) \KS(\to \pip \pim)$ decays collected in Run~2~\cite{LHCb-PAPER-2023-013}.
The considered final states are $\jpsi(\to \mun\mup)\KS$, $\phi(2S)(\to\mun\mup)\KS$ and $\jpsi(\to e^+ e^-)\KS$. 
Due to the small $\Bd$ mass-width difference, the $\CP$ asymmetry is approximately $\mathcal{A}^{\CP}(t) \approx S \sin(\Delta m_d t) - C \cos(\Delta m_d t)$, where $S$ is equal to $\sin(2\beta)$ in the SM, besides CKM-suppressed contributions from penguin topologies.
The determination of the $\CP$ observables requires a flavour-tagged and time-dependent analysis, resulting in $S_{\psi \KS} = 0.717 \pm 0.013 \pm 0.008$ and $C_{\psi \KS} = 0.008 \pm 0.012 \pm 0.003$.
Figure~\ref{fig:PsiKs_ACP} shows the measured $\CP$ asymmetry as a function of $\Bd$ decay time.

\subsection{Measurement of \texorpdfstring{$\Delta \Gamma_s$}{DeltaGammas}}
The measurements of the $\Bs$ decay-width difference $\Delta \Gamma_s$ using $\Bs\to\jpsi\phi$ decays show a tension between the LHC experiments~\cite{HFLAV}.
Consequently, performing measurements using other decay modes is important.
As such, $\Delta \Gamma_s$ is determined as the decay-with difference between $\CP$-even $\Bs\to\jpsi\eta^\prime$ and mostly $\CP$-odd $\Bs\to\jpsi\pim\pip$ decays, collected between 2011 and 2018~\cite{LHCb-PAPER-2023-025}.
The yield ratio of these decays in bins of decay time depends on $\DGs$, allowing for a $\chisq$ fit to obtain $\DGs$.
Finally, the value of $\Delta \Gamma_s = 0.087 \pm 0.012 \pm 0.009 \ps^{-1}$ is reported, which is compatible with the HFLAV average.

\section{Charm measurements}
$\CP$ violation in $D$ mesons has been observed for the first time in 2019~\cite{LHCb-PAPER-2019-006}.
The compatibility of this result with SM is debated, therefore new measurements of $\CP$ violation in charm meson decays are crucial to clarify the picture~\cite{charm-anomaly}.
These proceedings discuss the time-dependent study of charm mixing and $\CP$ violation in $\Dz\to\Kp\pim$ decays and the usage of the energy-test method to study $\CP$ violation in the $\Dz\to\KS\Km\pip$ and $\Dz\to\pim\pip\piz$ decays.
The neutral $D$ meson flavour at production in these measurements is determined by requiring it to originate from $\Dstarp\to\Dz\pip$ decays.

\subsection{Charm mixing and \texorpdfstring{$\CP$}{CP} violation with \texorpdfstring{$\Dz\to\Kp\pim$}{D0->K+pi-}}
In this measurement, the Run 2 dataset is exploited to probe mixing and $\CP$-violating parameters by measuring the decay-time dependency of the ratio of wrong sign (WS) $\Dz\to\Kp\pim$ and right sign (RS) $\Dzb\to\Kp\pim$ decays, $\RKpip(t)$, and the ratio of $\CP$-conjugated processes, $\RKpim(t)$~\cite{LHCb-PAPER-2024-008}. 
The ratios are expanded as $\RKpipm(t) \approx \RKpi(1 \pm \AKpi) + \sqrt{\RKpi(1 \pm \AKpi)} (\cKpi\pm\DcKpi)(t/\tauDz) + (\cKpipr\pm\DcKpipr) (t/\tauDz)^2$, where $\cKpi$ and $\cKpipr$ are sensitive to mixing parameters, while \AKpi, \DcKpi and \DcKpipr are sensitive $\CP$ violation in the decay, the mixing and their interference, respectively.
The WS-to-RS yield ratio in 18 bins of $\Dz$ decay time is used to extract $R_{K\pi} = (343.1\pm2.0)\times 10^{-5}$, $c_{K\pi} =(51.4\pm3.5)\times 10^{-4}$ and $c^\prime_{K\pi} =(13\pm4)\times 10^{-6}$.
No evidence for $\CP$ violation is found.

\subsection{Search for \texorpdfstring{\CP}{CP} violation using the energy-test method}
A search for $\CP$ violation is conducted in the phase space of $\Dz\to\KS \Kpm\pimp$ decays~\cite{LHCb-PAPER-2023-019} and $\Dz\to\pim\pip\piz$ decays~\cite{LHCb-PAPER-2023-005} using the energy-test method.
This method studies local $\CP$ violation by checking if the phase spaces of $\Dz$ and $\Dzb$ decays follow the same underlying distribution.
A test statistic $T$ is measured using the data which is compared with the $T$ distribution under the null hypothesis.
The resulting p-values correspond to $70\%$ for $\Dz\to\KS\Km\pip$, $66\%$ for $\Dz\to\KS\Kp\pim$ and $62\%$ for $\Dz\to\pim\pip\piz$; showing no evidence for $\CP$ violation.

\section{Conclusions}
Great progress is made in both beauty and charm measurements in LHCb.
The experimental uncertainty on $\gamma$ has dropped 3 degrees, with the latest HFLAV average providing \mbox{$\gamma = (66.5^{+2.8}_{-2.9})\degrees$~\cite{HFLAV}}, whereas the indirect determination provides $\gamma = (66.3^{+0.7}_{-1.9})\degrees$~\cite{CKMFitter}.
The sensitivity gap between direct and indirect determinations of $\gamma$ is narrowing, and closing this gap is one of the physics goals of LHCb.
Additionally, the most precise determinations of the CKM angles $\beta$ and $\beta_s$ are discussed.
Finally, mixing and $\CP$ violation in charm mesons is probed, further constraining the charm-mixing parameters and finding agreement with $\CP$ symmetry and the SM.

\section*{References}

{\scriptsize
\bibliography{ButterJordy}
}

\end{document}